\documentclass[twocolumn]{aastex631}

\usepackage[utf8]{inputenc}
\usepackage{graphicx}
\usepackage{amsmath}
\usepackage{float}  
\usepackage{subfigure}
\usepackage{booktabs} 
\usepackage{threeparttable}
\usepackage{array}
\usepackage{hyperref}
\usepackage{tabularx}

\newcommand\feh{{\rm [Fe/H]}}
\newcommand\logg{{\rm log}\,g}
\newcommand\teff{T_{\rm eff}}

\graphicspath{{figures/}}

\shorttitle{Empirical extinction coefficients for the \textit{Swift}-UVOT passbands}
\shortauthors{Fang et al.}

\begin{document}

\title{Empirical extinction coefficients for the \textit{Swift}-UVOT optical-through-ultraviolet passbands}

\author{Fang Yi}
\affil{Institute for Frontiers in Astronomy and Astrophysics, Beijing Normal University,  Beijing 102206, China}
\affil{Department of Astronomy, Beijing Normal University, Beijing, 100875, P.R.China \\}

\author{Yuan Haibo}
\affil{Institute for Frontiers in Astronomy and Astrophysics, Beijing Normal University,  Beijing 102206, China}
\affil{Department of Astronomy, Beijing Normal University, Beijing, 100875, P.R.China \\}

\author{Zhang Ruoyi}
\affil{Institute for Frontiers in Astronomy and Astrophysics, Beijing Normal University,  Beijing 102206, China}
\affil{Department of Astronomy, Beijing Normal University, Beijing, 100875, P.R.China \\}

\author{Gao Jian}
\affil{Institute for Frontiers in Astronomy and Astrophysics, Beijing Normal University,  Beijing 102206, China}
\affil{Department of Astronomy, Beijing Normal University, Beijing, 100875, P.R.China \\}

\author{Xu Shuai}
\affil{Institute for Frontiers in Astronomy and Astrophysics, Beijing Normal University,  Beijing 102206, China}
\affil{Department of Astronomy, Beijing Normal University, Beijing, 100875, P.R.China \\}

\correspondingauthor{Haibo Yuan email:yuanhb@bnu.edu.cn}

\begin{abstract}
We calculated empirical reddening and extinction coefficients with respect to the dust reddening map of Schlegel et al. for the \textit{Swift}-UVOT passbands, using the "star pair" method and 
photometric data from the UVOT Serendipitous Source Catalogue and 2MASS as well as spectroscopic data from LAMOST Data Release 7.
The reddening coefficients for the $UVW2-UVM2$, $UVM2-UVW1$, $UVW1-U$, $U-B$, and $B-V$ colors are $-1.39$, 2.08, 0.78, 0.72, and 0.84, respectively. The extinction coefficients for the $UVW2$, $UVM2$, $UVW1$, $U$, $B$, and $V$ bands are 5.60, 6.99, 4.91, 4.13, 3.41 and 2.57, respectively.
The numbers are consistent with predictions by the
Fitzpatrick's extinction law of $R(V)$ = 3.0. 
Temperature-dependent variations of the coefficients are found and discussed, particularly for the ultraviolet passbands. 
We recommend using the new reddening and extinction coefficients in future when dereddening the \textit{Swift}-UVOT data.

\end{abstract}

\section{Introduction} \label{sec:intro}

Dust grains are prevalent throughout the universe and cause extinction and reddening of stellar light from the ultraviolet (UV) to the infrared (IR). Reddening maps are usually used along with reddening coefficients to correct the reddening in different colors, and reveal astronomical objects' intrinsic spectral energy distribution and properties. \cite{1998ApJ...500..525S} provided one of the most widely used reddening maps (SFD map hereafter). 
For a certain passband $a$, we have: $A(a) = R(a) * E (B-V)$, where $A(a)$ is the total extinction in the band $a$, $E (B-V)$ is the color excess in the $B-V$ color, and $R(a)$ is the extinction coefficient of band $a$. The reddening coefficient of $a-b$ color can be expressed by $R (a-b) = E(a-b)/E(B-V) = R(a)-R(b)$. 
Values of $R(a)$ and $R(a-b)$ can be predicted theoretically by a given extinction curve (e.g., \citealp{1989ApJ...345..245C}, \citealp{1994ApJ...422..158O}, \citealp{1999PASP..111...63F}) or determined empirically by observational data (e.g., \citealp{2011ApJ...737..103S}, \citealp{2013MNRAS.430.2188Y}, \citealp{2018ApJ...861..153S},
\citealp{2023ApJS..264...14Z}
).

Studying extinction in the UV is particularly important. On one hand, extinction coefficients in the UV can place much stronger constraints on extinction curves than those in the optical and IR. On the other hand,  
the extinction in the UV is much larger than that in the visible, thus serving as a better tracer of dust.

\begin{figure}[htbp]
        \centering
        \vspace{0.2cm}
        \includegraphics[width=8cm]{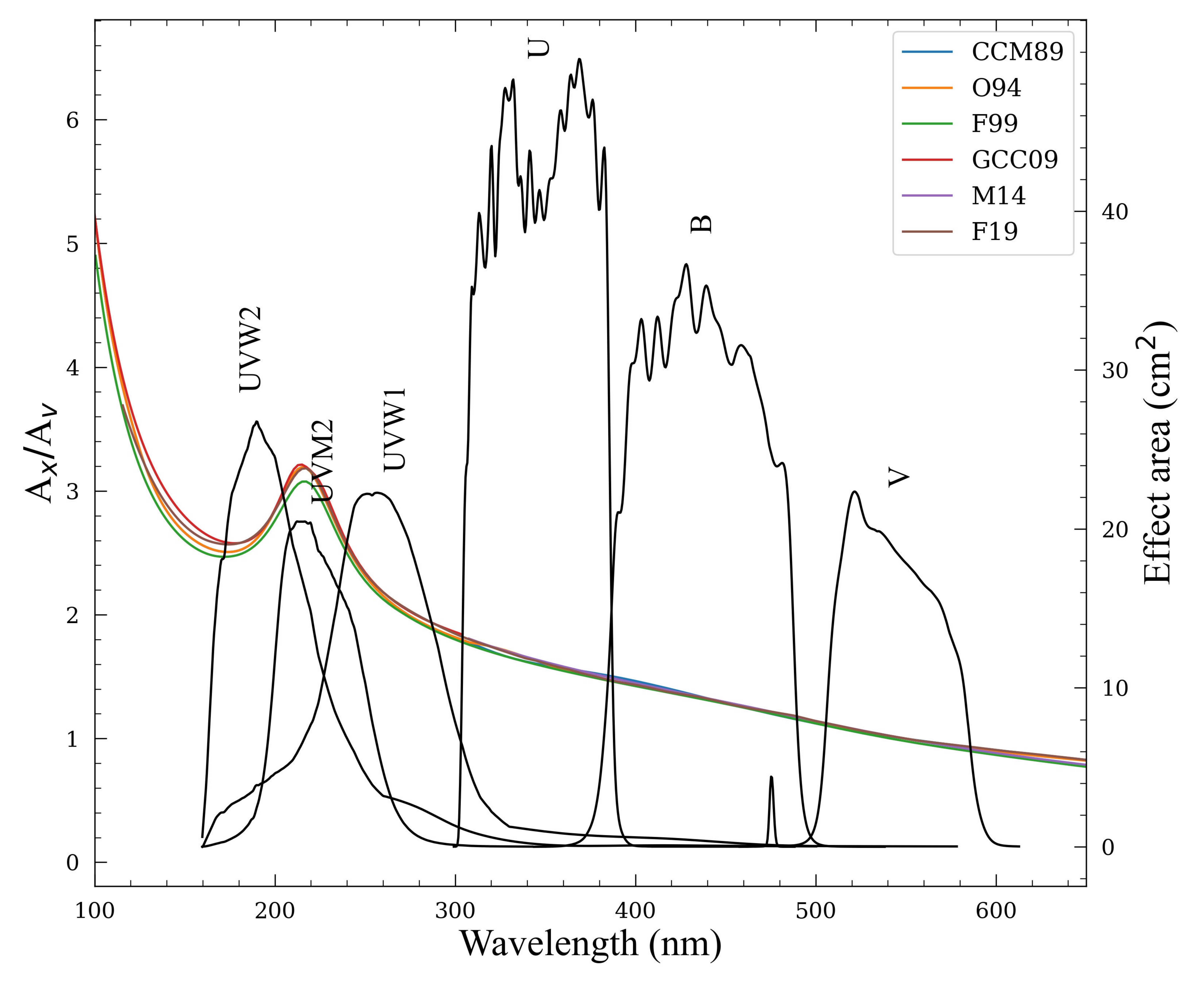}
        \caption{
        Response functions of the \textit{swift}-UVOT's $UVW2$, $UVM2$, $UVW1$, $U$, $B$, and $V$ filters \citep{2011AIPC.1358..373B}. A series of different dust extinction curves are also over-plotted.  References: \citealp{1989ApJ...345..245C}\,(CCM89), \citealp{1994ApJ...422..158O}\,(O94), \citealp{1999PASP..111...63F}\,(F99), \citealp{2009ApJ...705.1320G}\,(GCC09), \citealp{2014A&A...564A..63M}\,(M14),  \citealp{2019ApJ...886..108F}\,(F19).
 }
        \label{fig:1}
    \end{figure}

Since the Earth's atmosphere blocks almost all UV radiation, space telescopes are needed to receive UV light. The Neil Gehrels Swift Observatory (formerly known as the \textit{Swift} satellite, \citealp{2004ApJ...611.1005G}) is a NASA medium explorer designed to study gamma-ray bursts. It has three telescopes: the Burst Alert Telescope (\citealp{2005SSRv..120..143B}), the X-Ray Telescope (\citealp{2005SSRv..120..165B}) and the Ultraviolet/Optical Telescope (UVOT; \citealp{2000SPIE.4140...76R,2005SSRv..120...95R}). UVOT is a 30 cm telescope with seven broadband filters and two grisms that observes over a wavelength range of 1650--6000Å. Its three NUV filters cover the range of the well-known 2175Å bump (see Figure \ref{fig:1}) and can be used to constrain the shape of different UV extinction curves.

In this work, we will determine empirical reddening and extinction coefficients with respect to the SFD dust map for the \textit{Swift}-UVOT passbands.
The paper is structured as follows: Section \ref{sec:data and method} introduces the data and method we used and describes the data selection for target and control samples. Empirical reddening and extinction coefficients for the UVOT passbands are presented, compared and discussed in Section \ref{third:style}. We summarize in Section \ref{forth:style}.

\section{Data and method} \label{sec:data and method}

\subsection{Data}
\label{2.1}
 We used photometric data from the UVOT Serendipitous Source Catalogue (UVOTSSC; \citealp{2014styd.confE..37P}) and spectroscopic data from LAMOST Data Release 7 (DR7; \citealp{2015RAA....15.1095L}).
 The UVOTSSC contains 6,200,016 sources which are detected by at least one of UVOT's six filters during its first five years of operation. Most of their UV sources are above the depth limit of the GALEX's all-sky imaging survey (\citealp{2014styd.confE..37P}).
LAMOST DR7 contains spectroscopic data for 6,199,917 stars, including derived effective temperature $T_{eff}$, surface gravity log \textit{g}, metallicity [Fe/H], and their uncertainties (\citealp{2015RAA....15.1095L}).
We also used data from the 2MASS Point Source Catalog (2MASSPSC; \citealp{2006AJ....131.1163S}).

    \begin{figure}[b]
        \centering
        \vspace{0.2cm}
        \includegraphics[width=9cm]{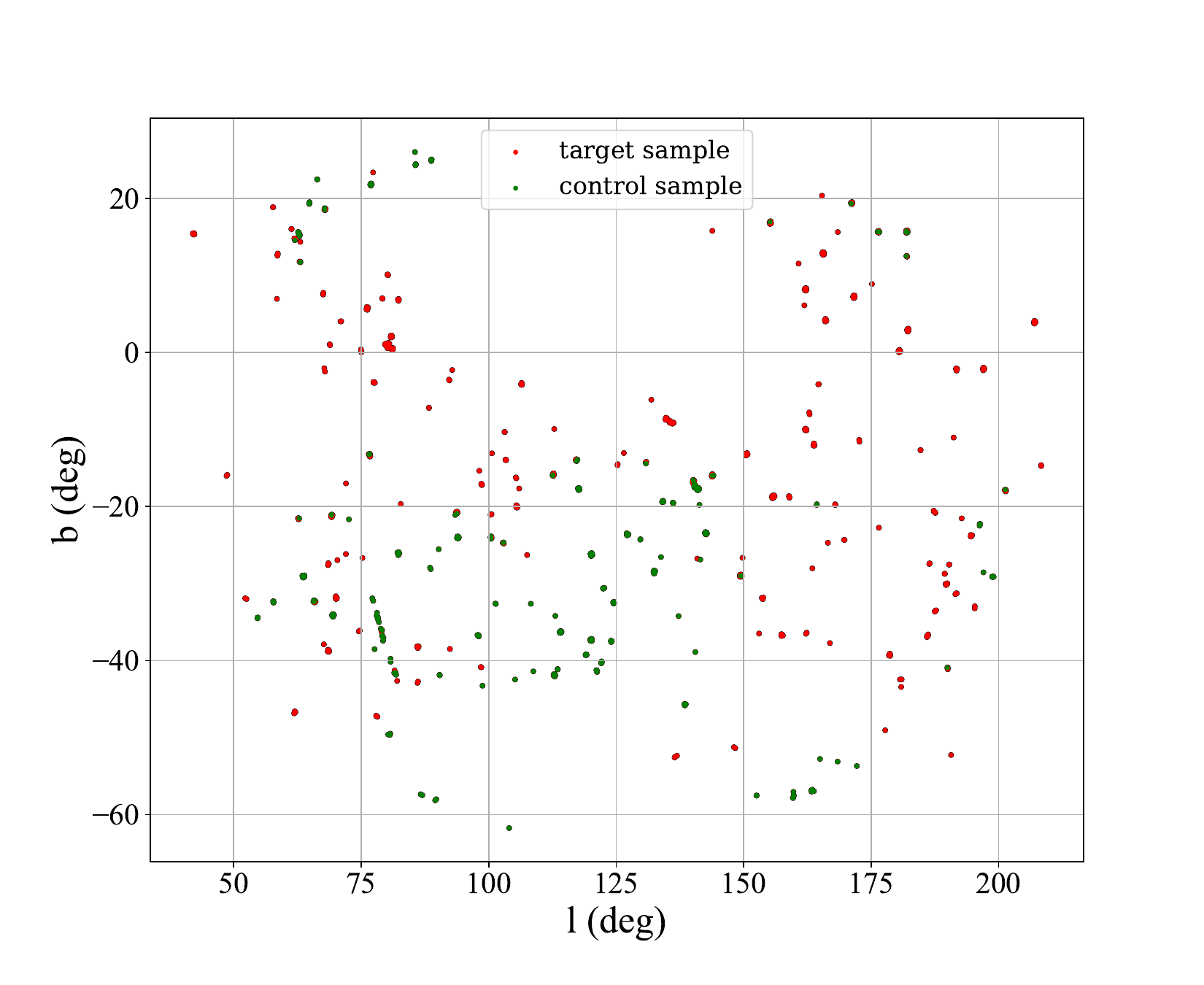}
        \caption{Spatial distribution of the  target (red dots) and control (green dots)  samples.}
        \label{fig2}
    \end{figure}

    \begin{figure}[t]
        \centering
        \vspace{0.2cm}
        \includegraphics[width=9cm]{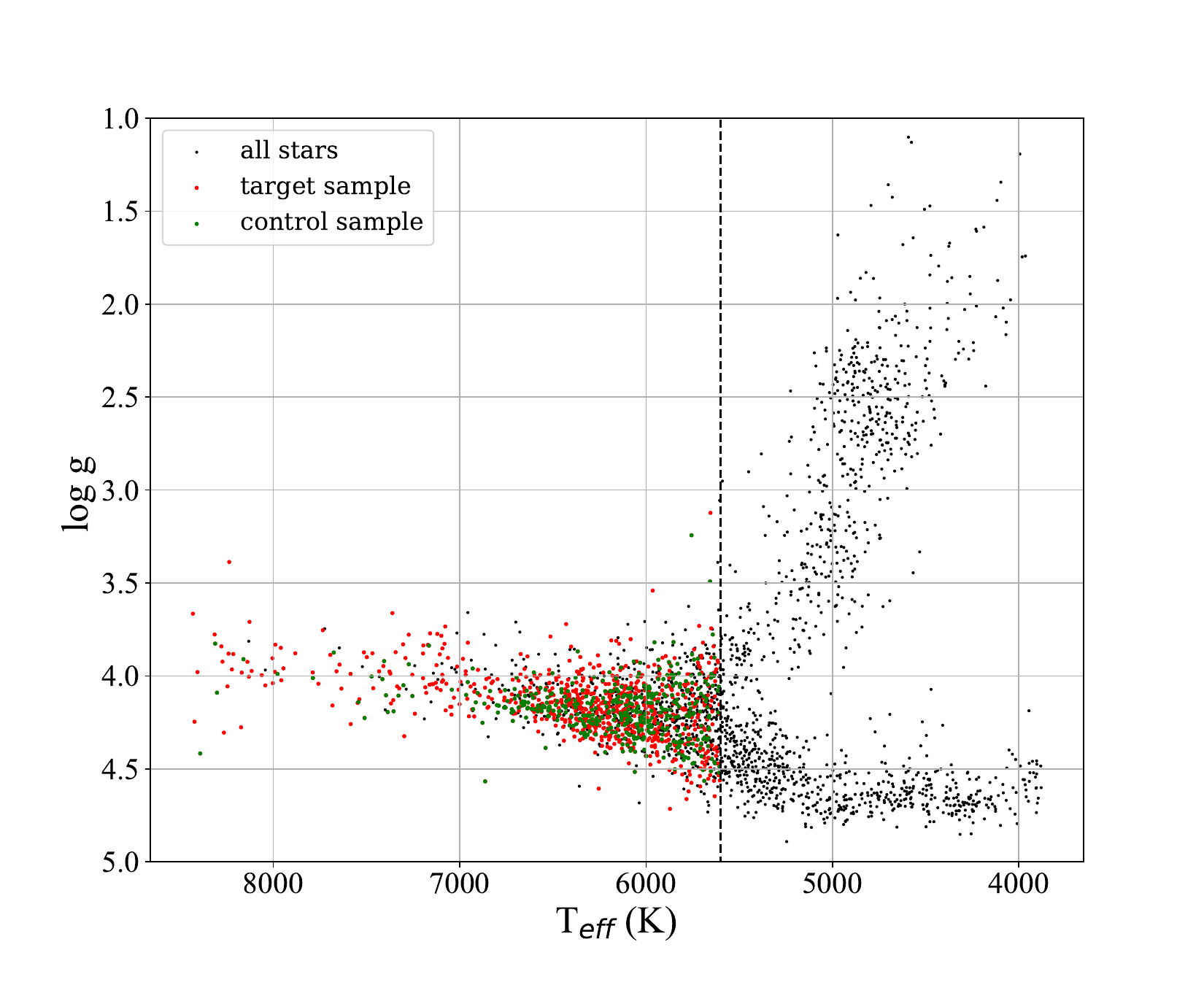}
        \caption{
        Distribution of target (red dots) and control (green dots) stars in the H-R diagram. Black points represent LAMOST-UVOT common stars. The vertical dashed line marks $T_{eff}$ = 5600\,K, which is to eliminate the relatively cool stars.}
        
        \label{fig3}
    \end{figure}

\subsection{Method}
·
We used the same standard pair method as \cite{2013MNRAS.430.2188Y}, which was first devised by \cite{1965ApJ...142.1683S}.
The method assumes that stars with the same stellar atmospheric parameters have the same intrinsic colors. Thus, the intrinsic colors of a reddened target star can be derived from its control counterparts with similar atmospheric parameters that have well-known extinction.
Meanwhile, we assumed that the intrinsic colors of the target and its control stars vary linearly with $T_{eff}$, log \textit{g}, and [Fe/H], within a certain range.
The method has been used in a number of studies (e.g., \citealp{2015MNRAS.448..855Y}, \citealp{2017MNRAS.467.1890X}, \citealp{2020ApJ...905L..20R}, \citealp{2022ApJS..260...17S}, \citealp{2023ApJS..264...14Z})

For target stars with $\teff$ lower than 7500K, control stars were selected from the control sample (see Section \ref{2.3}) such that their $\teff$, $\logg$, and $\feh$ differ from those of the target by no more than 300 K, 0.5 dex and 0.5 dex, respectively. For target stars with
$\teff$ higher than 7500\,K, their control stars were selected as those of $\Delta\teff < (0.1 \times \teff$)\,K, due to the small number of hot control stars available in this work.

Firstly, we  used an initial set of reddening coefficients and $E(B-V)$ values from SFD dust map to deredden the control sample and obtain the intrinsic colors. Note that the 
determination of final reddening coefficients was insensitive to the initial reddening coefficients, as the reddening values of the control sample were relatively small. Secondly, since we assumed that the intrinsic colors vary linearly with the $\teff$, $\logg$, and $\feh$ within a certain range, the intrinsic colors of the target sample can be derived using the fitting parameters for the control sample.
Thirdly, after obtaining the intrinsic colors of the target sample, a new set of reddening coefficients was derived by comparing the estimates of reddening relative to $E(B-V)$ for the target sample. Three-sigma clippings were performed during the linear fitting process to exclude outliers. Only few stars were excluded. The new set of reddening coefficients can be used to replace the initial reddening coefficient obtained in the first step.
Fourthly, iterations were performed until the newly derived set of reddening coefficients agreed with the ones used for dereddening the control sample.  After obtaining the reddening coefficients for two neighboring bands and assuming a predetermined value for the extinction coefficient of a specific passband, one can finally determine the extinction coefficients for all bands. In this work, we used the reddest passband of 2MASS as the specific passband, because the impact of dust in the infrared band is relatively small.

\subsection{Target sample and control sample}
\label{2.3}
We obtained our target sample by cross-matching the three datasets mentioned in Section\ref{2.1}. Several cuts were made to select the target stars.
First, we excluded stars with $\teff$ lower than 5600K. This is because cold stars show strong stellar activity in the UV (\citealp{2018ApJS..235...16B}), which could affect the measurements of reddening. Then we required the UVOT and 2MASS magnitude errors  smaller than 0.1 mag and the LAMOST spectral signal-to-noise ratio in the $g$ band larger than 20, in order to select reliable photometric and spectroscopic data. Upon cross-referencing with LAMOST DR7 database, we identified only 25766 common sources.  To ensure consistency in the subsequent analysis, we opted to impose the condition of data availability in all bands from UVOT.
A number of 1142 target stars were collected.

We obtained control sample from the target stars above by further requiring $E(B-V)_{SFD} < 0.075$. For stars with $\teff$ above 6500\,K, $E(B-V)_{SFD}$ was relaxed to less than 0.15. 
The control sample contained 425 stars. 

\begin{figure}[t]
        \centering
        \vspace{0.3cm}
        \includegraphics[width=9.5cm]{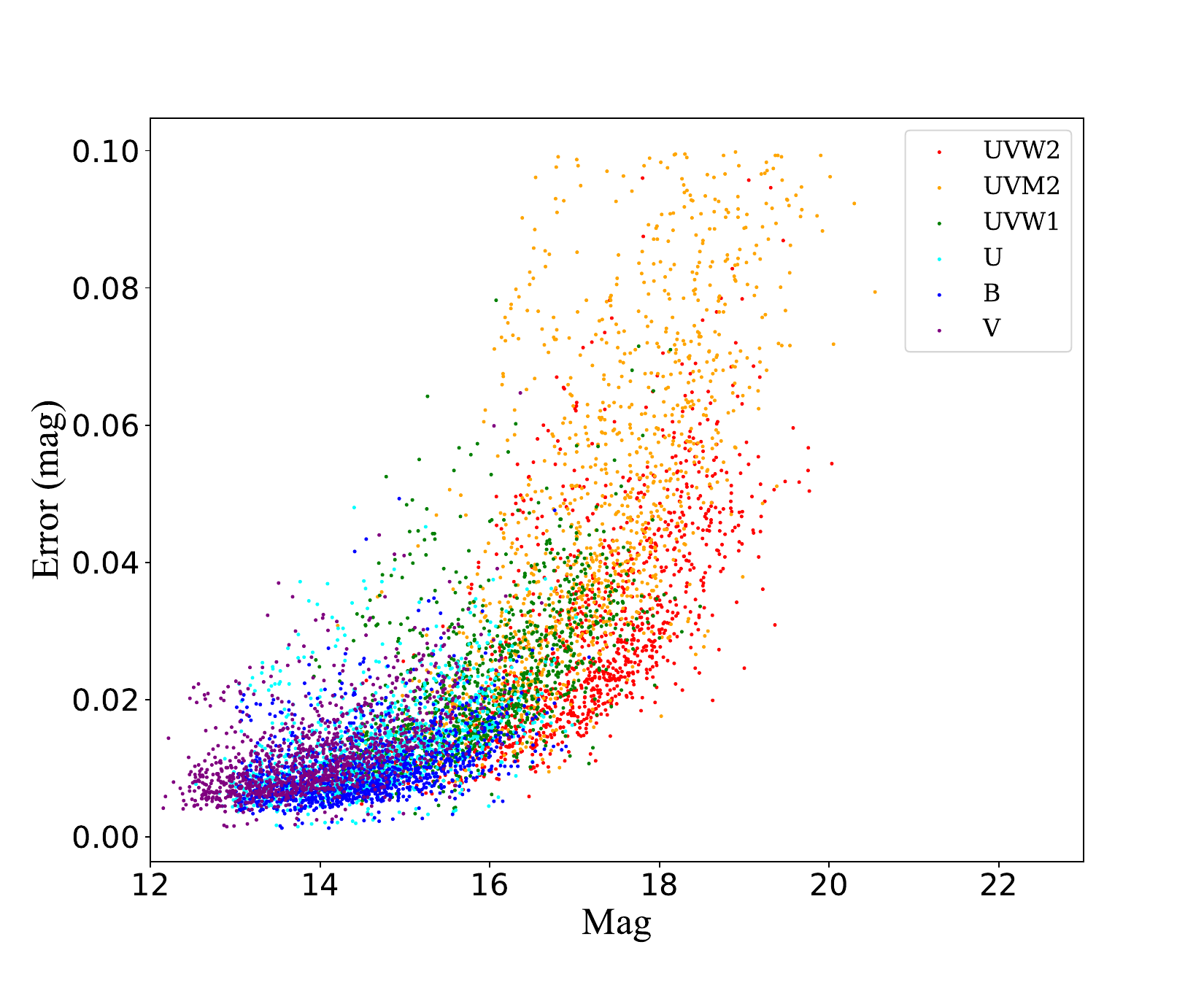}
        \caption{Photometric error as a function of magnitude for different UVOT passbands of target sample. }
        \label{fig4}
    \end{figure}

Figure \ref{fig2} plots the spatial distribution of the 
target and control stars. Most control stars are 
in the high Galactic latitude region. 
Figure \ref{fig3} shows the distribution of target and control stars in the LAMOST H-R diagram.
Figure \ref{fig4} illustrates photometric uncertainty as a function of magnitude of the six UVOT passbands for the target sample. For optical bands, 
most stars have magnitude errors  smaller than 0.02\,mag. For UV bands, most stars  have magnitude errors smaller than 0.05\,mag.

\section{Results and discussions} \label{third:style}
\subsection{Reddening and extinction coefficients}
\begin{figure*}
	\centering  
	\vspace{-0.1cm} 
	\subfigtopskip=5pt 
	\subfigbottomskip=5pt 
	\subfigcapskip=-4pt 
	\subfigure{
		\label{results}
		\includegraphics[width=0.4\linewidth]{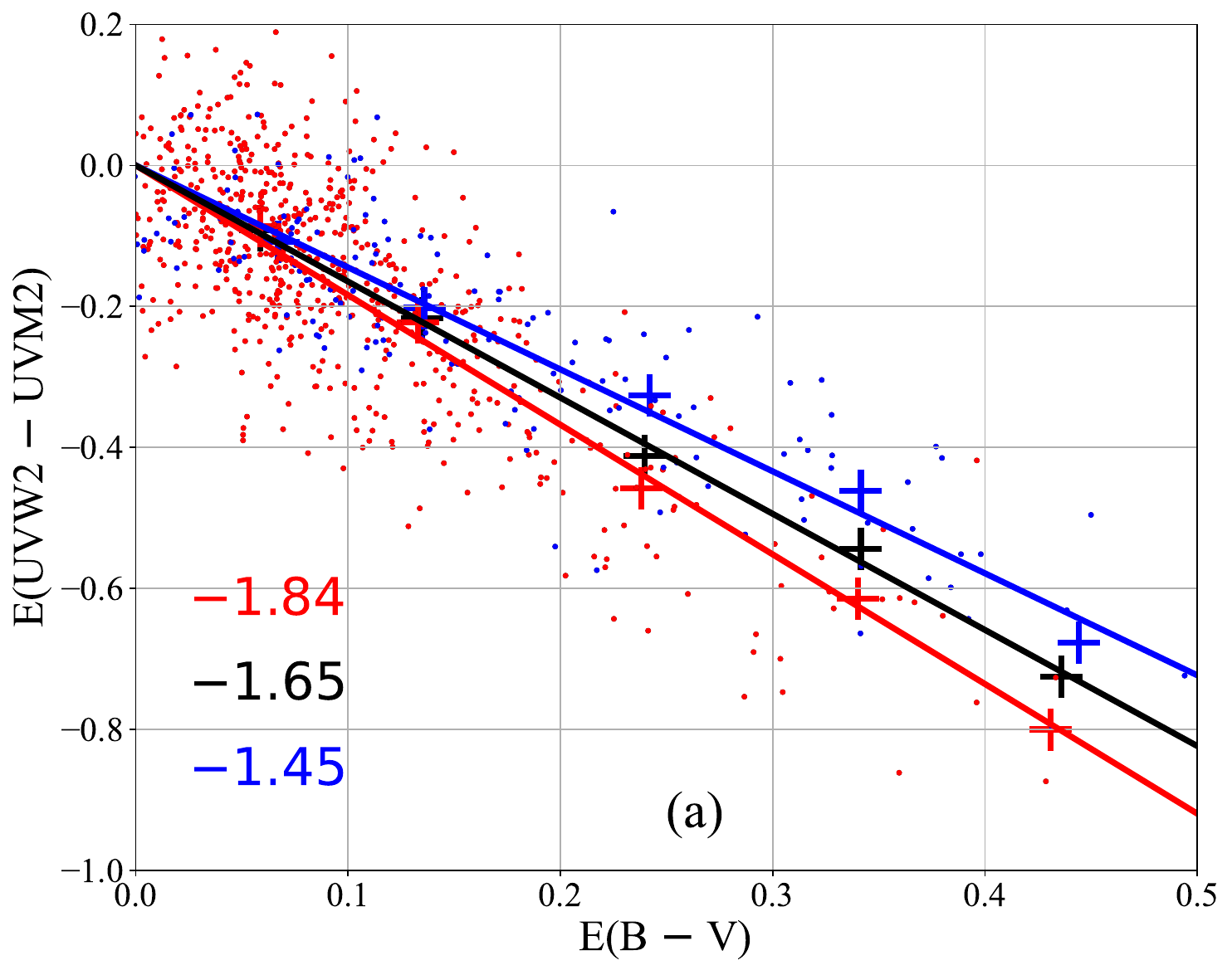}}
	\quad 
	\subfigure{
		\label{results}
		\includegraphics[width=0.4\linewidth]{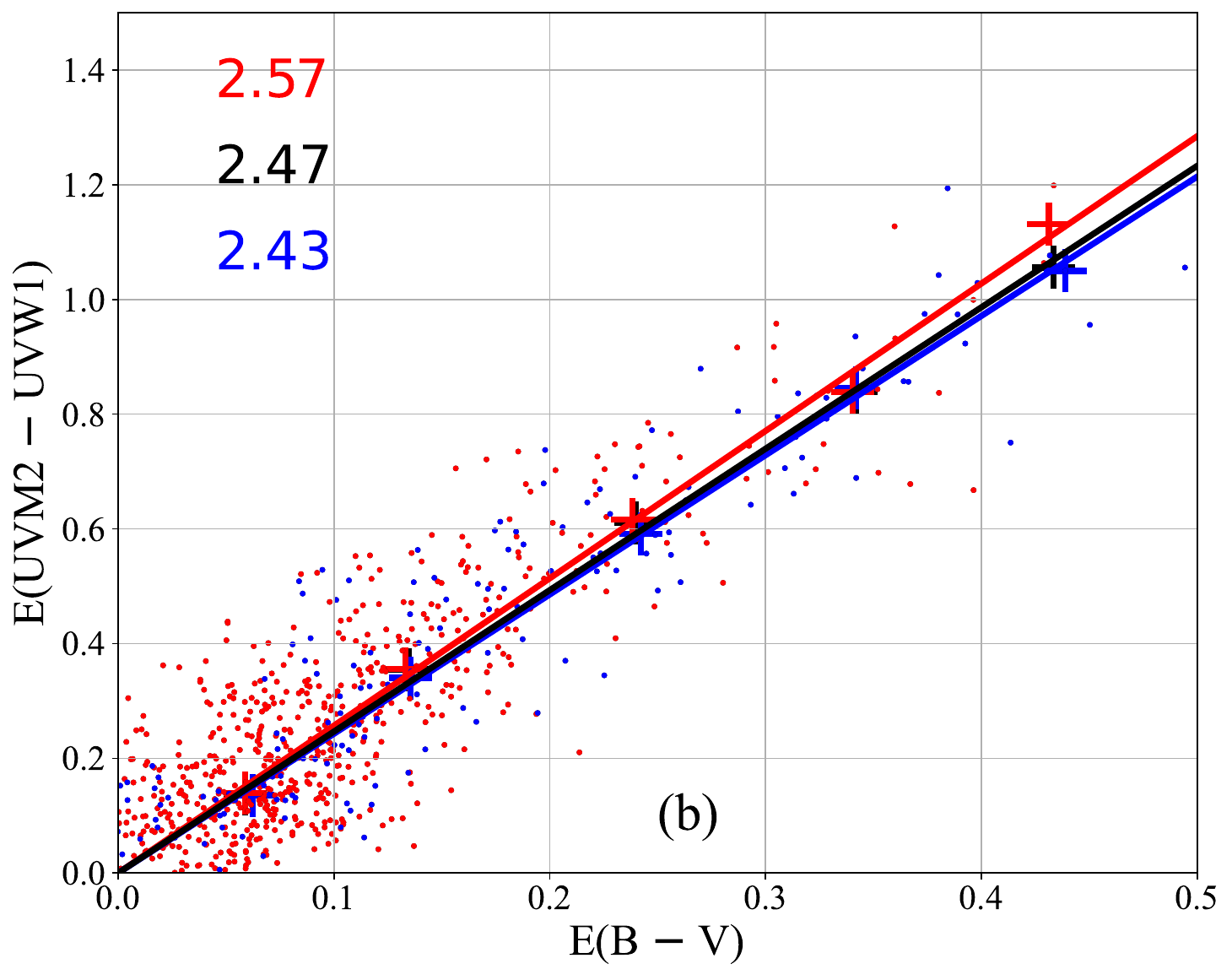}}
	\quad
	\subfigure{
		\label{results}
		\includegraphics[width=0.4\linewidth]{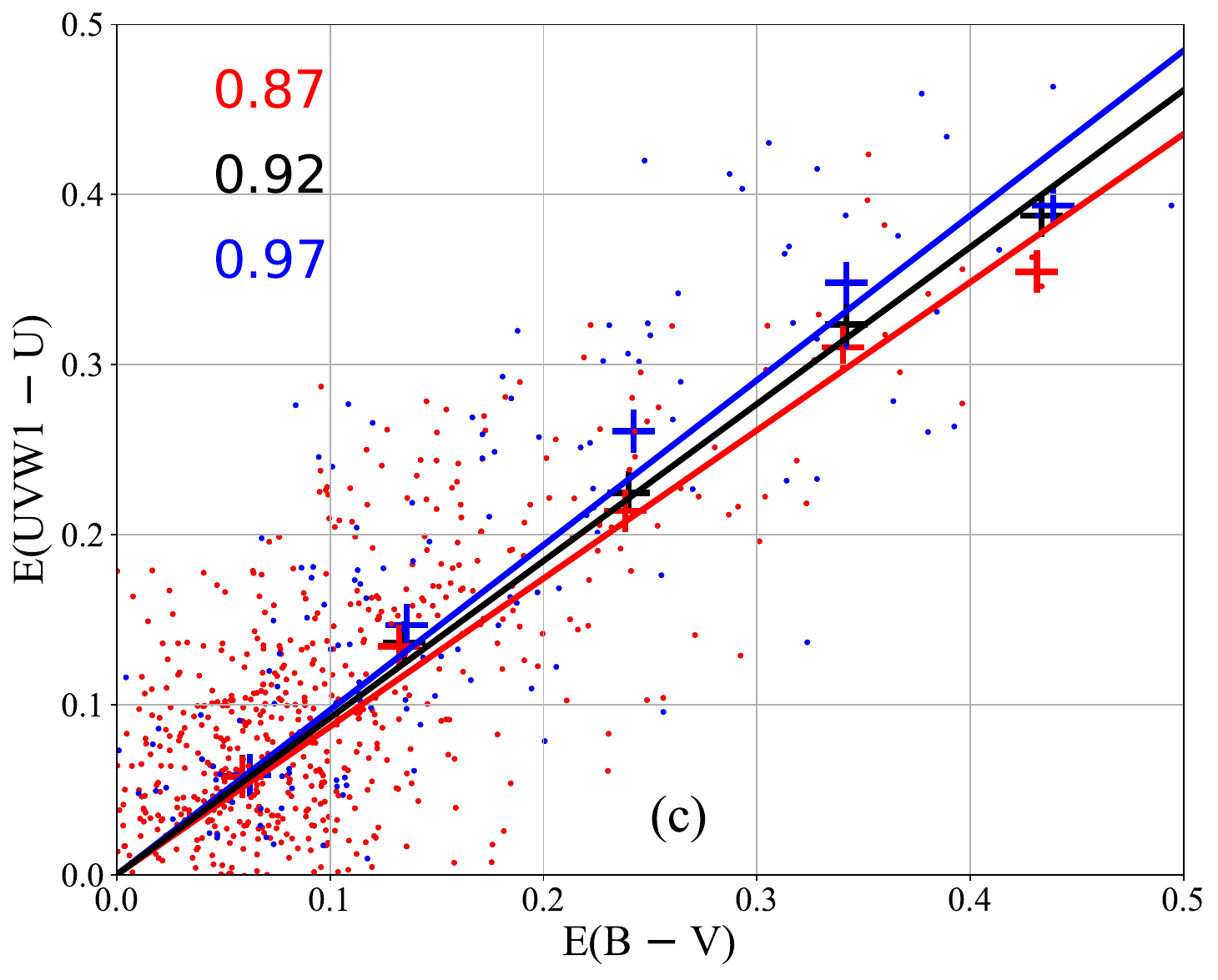}}
	\quad
	\subfigure{
		\label{results}
		\includegraphics[width=0.4\linewidth]{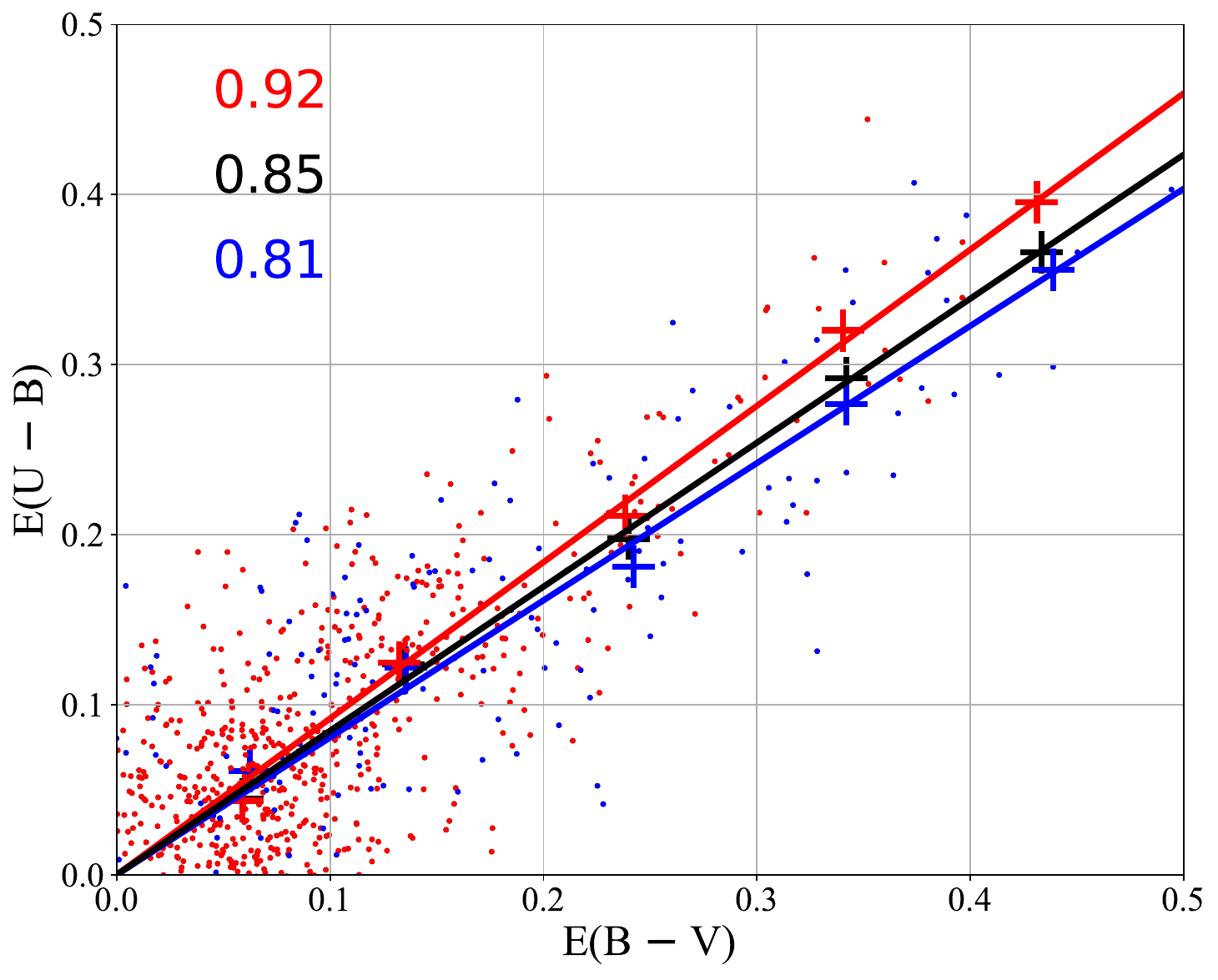}}
	\quad
	\subfigure{
		\label{results}
		\includegraphics[width=0.4\linewidth]{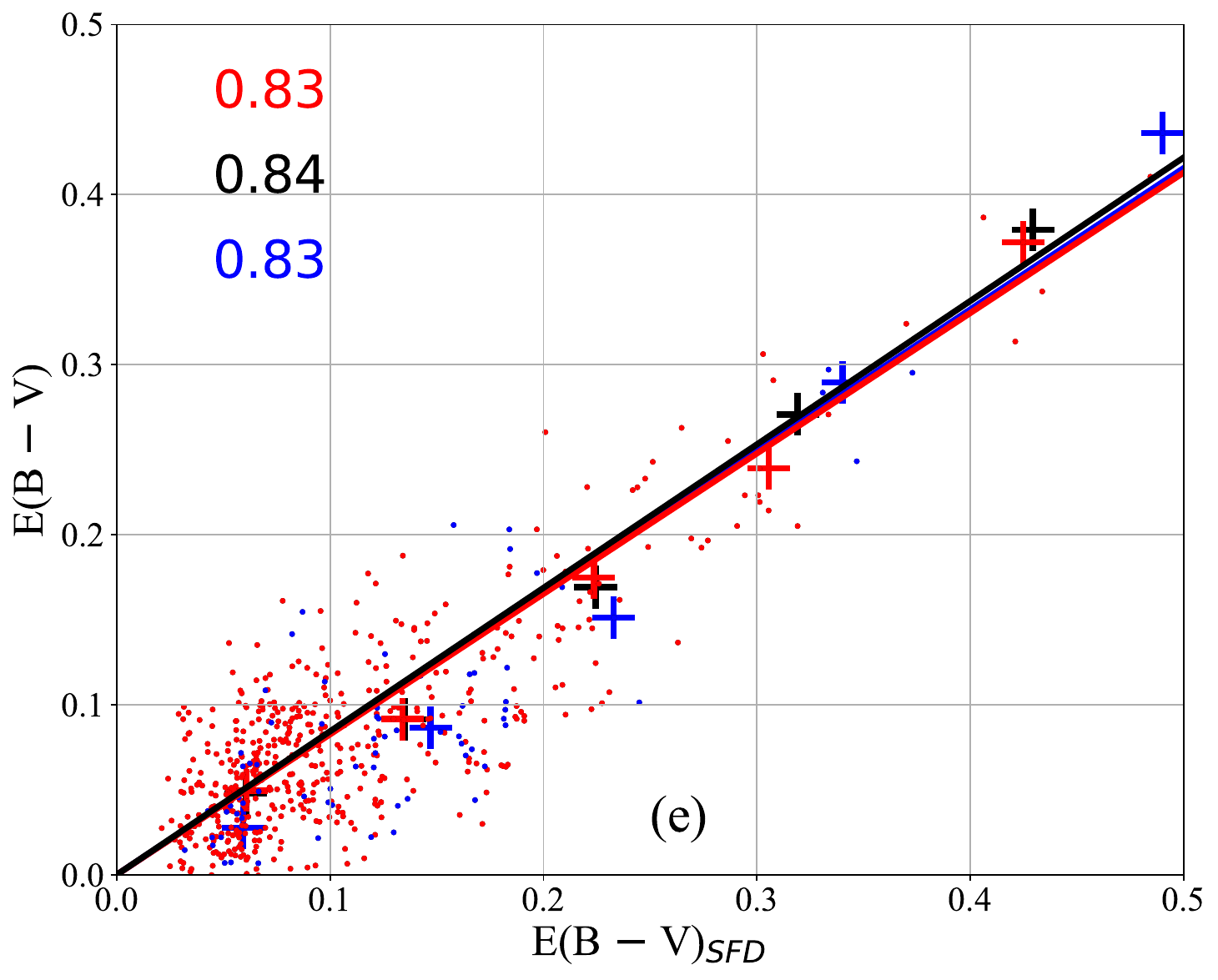}}
	\quad
	\caption{Reddening coefficients of the \textit{$UVW2-UVM2$, $UVM2-UVW1$, $UVW1-U$, $U-B$}, and \textit{$B-V$} colors deduced using the target sample. Red and blue dots represent stars of the low-temperature, high-temperature samples, respectively. The red, blue ,and black crosses represent median values for the low-temperature, high-temperature, and whole sample, respectively.  The lines show linear regression results. The slopes are also labeled.}
	\label{fig5}
\end{figure*}

Panels (a)--(d) in Figure \ref{fig5} display reddening values of $E(UVW2-UVM2)$, $E(UVM2-UVW1)$, $E(UVW1-U)$, and $E(U-B)$ versus reddening values of $E(B-V)$ from the target stars, respectively. 
For each panel, we binned the data points into five bins along the x-axis with a bin size of 0.1\,mag. The median values for each bin were calculated after 3-$\sigma$ clipping and are shown as black crosses.
Linear regressions were performed on these median values with equal weight for each point. The results are shown as black solid lines and forced to pass through the origin. The slopes, i.e., color excess ratios, are also 
labelled in black. Note that these results are independent of any systematic uncertainties in the SFD map.

To convert the above color excess ratios into reddening coefficients with respect to the SFD map, panel (e) compares $E(B-V)$ from UVOT with $E(B-V)_{SFD}$. To avoid systematic errors in the SFD map at low Galactic latitudes, we only use stars with $|b| > 10\degr$ in panel (e).
We obtained $R(B-V) = 0.84$. During the subsequent simulations using stellar atmospheric models (see Section\,\ref{3.2}), we discovered that the theoretical $R(B-V)$ value for UVOT is 0.96, corresponding to a 14\% ($0.96/0.84-1$) discrepancy.
The result is consistent with \cite{2010ApJ...725.1175S} and \cite{2013MNRAS.430.2188Y}, who found that the SFD map overestimates $E(B-V)$ values by 14\%.
Given $R(B-V) = 0.84$ and the color excess ratios in panels (a) -- (d), $R(UVW2-UVM2)$, $R(UVM2-UVW1)$, $R(UVW1-U)$, and $R(U-B)$ values with respect to the SFD map were calculated.  
The numbers are listed in the second column of 
Table \ref{table1}.

To calculate  extinction coefficients for the UVOT passbands, we first determined the values of 
$R(V-J)$, $R(J-H)$, and $R(H-K_s)$ using the same star-pair method. Then we adopted 
$R(K_s)$ = 0.3 as the reference point.
This value was derived using Fitzpatrick's law for $R(V)$=3.0, assuming that the SFD map overestimates $E(B-V)$ by 14\%.
The F99 extinction law of $R(V)$=3.0 was used because it is consistent with our data (see Section\ref{forth:style}). The extinction coefficients for the UVOT passbands are given in Table \ref{table2}, and these coefficients in Table \ref{table1} and Table \ref{table2} have already taken into account the 14\% overestimation.

In Table 1 of \cite{2019MNRAS.486..743D}, the ratios of extinction in the $UVW2$, $UVM2$, and $UVW1$ bands to that in the $V$ band ($A_\lambda$/$A_V$ ), are given as 2.67, 2.80, and 2.26, respectively. Our study yields values of 2.18, 2.71, and 1.91 for the corresponding bands. It is not surprising that the results obtained by the two methods differ, as this discrepancy arises from the different approaches employed. \cite{2019MNRAS.486..743D} convolved the CCM89 extinction curve with the transmission curve of the filters, assuming a flat response curve within each passband but this assumption is invalid for stars.

Note that the errors of reddening/extinction coefficients given in Table \ref{table1} and Table \ref{table2} are just formal errors. They are very likely under-estimated,
particularly for the UV bands, as the shape of UV extinction may vary strongly along different lines of sight (e.g., \citealp{1985ApJS...59..397S};   \citealp{2019ApJ...886..108F}). Such effect was ignored due to the limited number of stars available in this work.

\begin{table}[!htbp]
\begin{minipage}[]{90mm}
\centering
\caption{\textit{R(a-b)} for various colors with respect to the SFD map.  To do reddening correction, these numbers should be directly multiplied by the SFD map.}\label{table1}
\setlength{\tabcolsep}{6mm}{
    \begin{tabular}{lll}
\hline
Color       & This work        & F99$^1$ \\ \hline
$UVW2-UVM2$ & $-1.39 \pm 0.02$ & $-$1.36 \\
$UVM2-UVW1$ & $2.08 \pm 0.05$  & 2.21    \\
$UVW1-U$    & $0.78 \pm 0.01$  & 0.66    \\
$U-B$       & $0.72 \pm 0.01$  & 0.73    \\
$B-V$       & $0.84 \pm 0.02$  & 0.84    \\
$V-J$       & $1.89 \pm 0.04$  & 1.93    \\
$J-H$       & $0.22 \pm 0.02$  & 0.25    \\
$H-K_s$     & $0.16 \pm 0.02$  & 0.14    \\ \hline
\end{tabular}}
    \begin{description}

        \item[$^1$] Prediction by 
        the F99 extinction law of \textit{$R(V)=3.0$} for a 6250\,K source spectrum at $E(B-V)=0.2$,
        assuming that SFD overestimates \textit{$E(B - V)$} by 14 per cent.

    \end{description}
\end{minipage}
\end{table}

\begin{table}[!htbp]
\centering
\caption{\textit{R(a)} for various passbands with respect to the SFD map.  To do extinction correction, these numbers should be directly multiplied by the SFD map.}\label{table2}
\setlength{\tabcolsep}{8mm}{
    \begin{tabular}{lll}
\hline
Passband & This work$^1$   & F99$^2$ \\ \hline
$UVW2$   & $5.60\pm 0.02$  & 5.70    \\
$UVM2$   & $6.99 \pm 0.05$ & 7.06    \\
$UVW1$   & $4.91 \pm 0.01$ & 4.85    \\
$U$      & $4.13 \pm 0.01$ & 4.19    \\
$B$      & $3.41 \pm 0.02$ & 3.46    \\
$V$      & $2.57 \pm 0.04$ & 2.62    \\
$J$      & $0.68 \pm 0.02$ & 0.68    \\
$H$      & $0.46 \pm 0.02$ & 0.46    \\
$K_s$    & $0.30 $         & 0.30    \\ \hline
\end{tabular}}
    \begin{description}
        \item[$^1$] Calculated using R(K$_s$) = 0.3 and reddening coefficients from Table\,1.
        \item[$^2$] Prediction of the 
         F99 extinction law of \textit{$R(V)=3.0$} for a 6250\,K source spectrum at $E(B-V)=0.2$, assuming that SFD overpredicts the true values of $E(B-V)$ by 14 per cent.
        
    \end{description}
\end{table}


\subsection{Comparison with the F99 extinction law}\label{3.2}
To compare the empirically measured reddening/extinction coefficients with theoretical values, we estimated predictions using the F99 extinction law and the BOSZ stellar atmosphere model (\citealp{2017AJ....153..234B}). The F99 extinction law was chosen because it matches well with observations (e.g., \citealp{2011ApJ...737..103S}; \citealp{2013MNRAS.430.2188Y}). 
When estimating theoretical values,
we selected typical an F-type dwarf with $\logg$ = 4.0, [Fe/H]=0, and $\teff$ = 6500\,K to represent our target sample, we adopted a typical reddening value of 0.2, and we assumed a 14\% overestimation of $E(B-V)$ by the SFD map.
Predictions of the F99 extinction law at different \textit{$R(V)$} values were calculated and compared with observations. 

We selected the $R(V)$ range between 2.5 and 3.5, with an interval of 0.1 for comparison. We calculated the sum of squared differences between the coefficients obtained from the model's calculations and the coefficients obtained in our work. The $R(V)$ value corresponding to the minimum sum of squared differences was considered as the optimal fitting $R(V)$ value. It was found that the F99 extinction law of \textit{$R(V)=3.0$} agrees with the measured data best. 
The results are shown in Figure\,6 and Figure \ref{fig7}.
One can see that the agreement is good, with max difference less than 0.05 in the optical and infrared and less than 0.12 in the UV.
These predicted reddening and extinction coefficients are listed in Tables \ref{table1} and \ref{table2}, respectively.

We found $R(V)$ = 3.0 agrees our result best instead of  $R(V)$ = 3.1. 
There are two main reasons for this: 1) relatively large measurement errors due to limited sample in this work and 2) spatial variations of $R(V)$.  \cite{2016ApJ...821...78S} suggested that $\sigma(R(V)) = 0.18$. Zhang et al. (2023, submitted) find that  $\sigma(R(V)) = 0.25$.

    \begin{figure}
        \centering
        \vspace{0.2cm}
        \includegraphics[width=9cm]{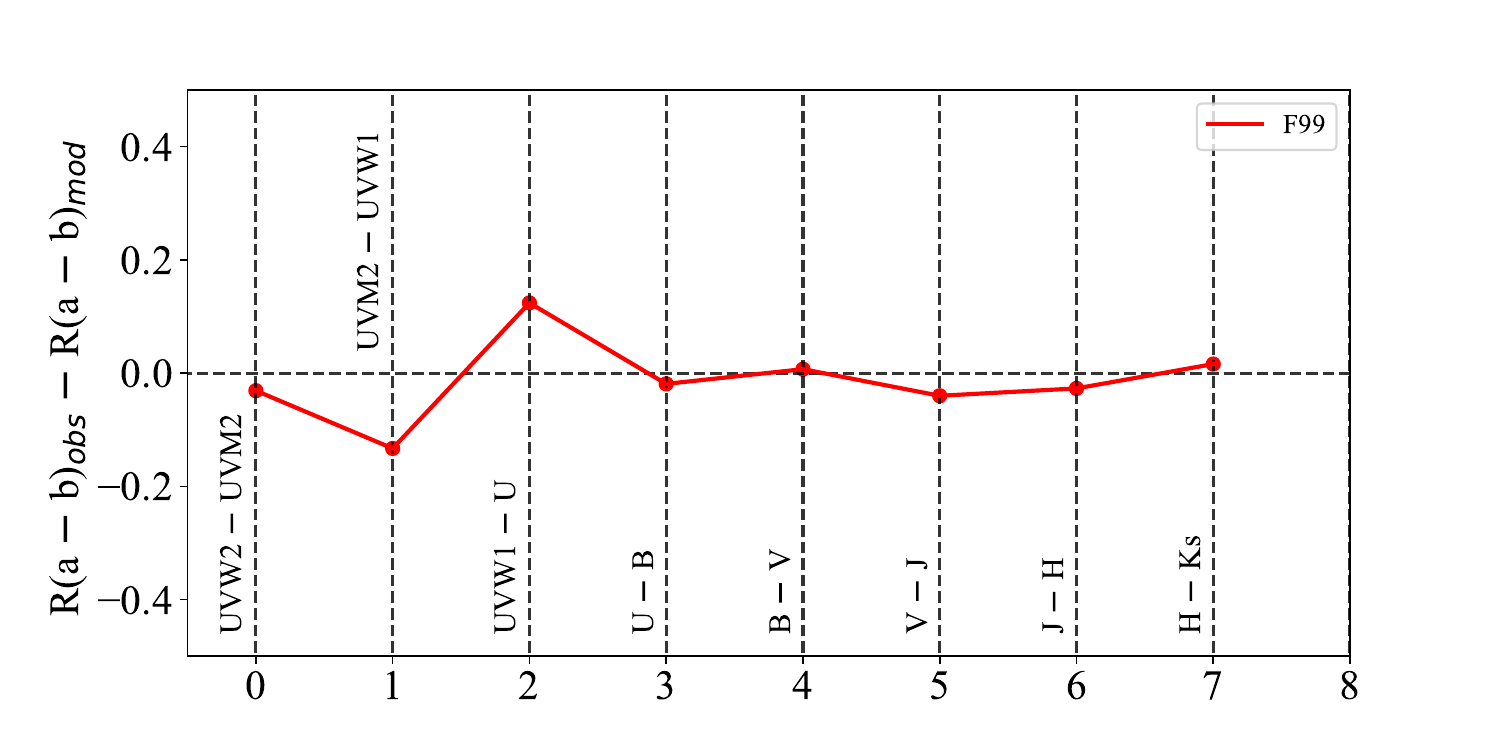}
        \caption{Comparison of the measured and 
        predicted reddening coefficients by the 
         F99 extinction law of \textit{$R(V)=3.0$} for a 6250\,K source spectrum at $E(B-V)=0.2$, assuming that SFD overestimates the true values of $E(B-V)$ by 14 per cent
        .}
        \label{fig6}
    \end{figure}

     \begin{figure}
        \centering
        \vspace{0.5cm}
        \includegraphics[width=9cm]{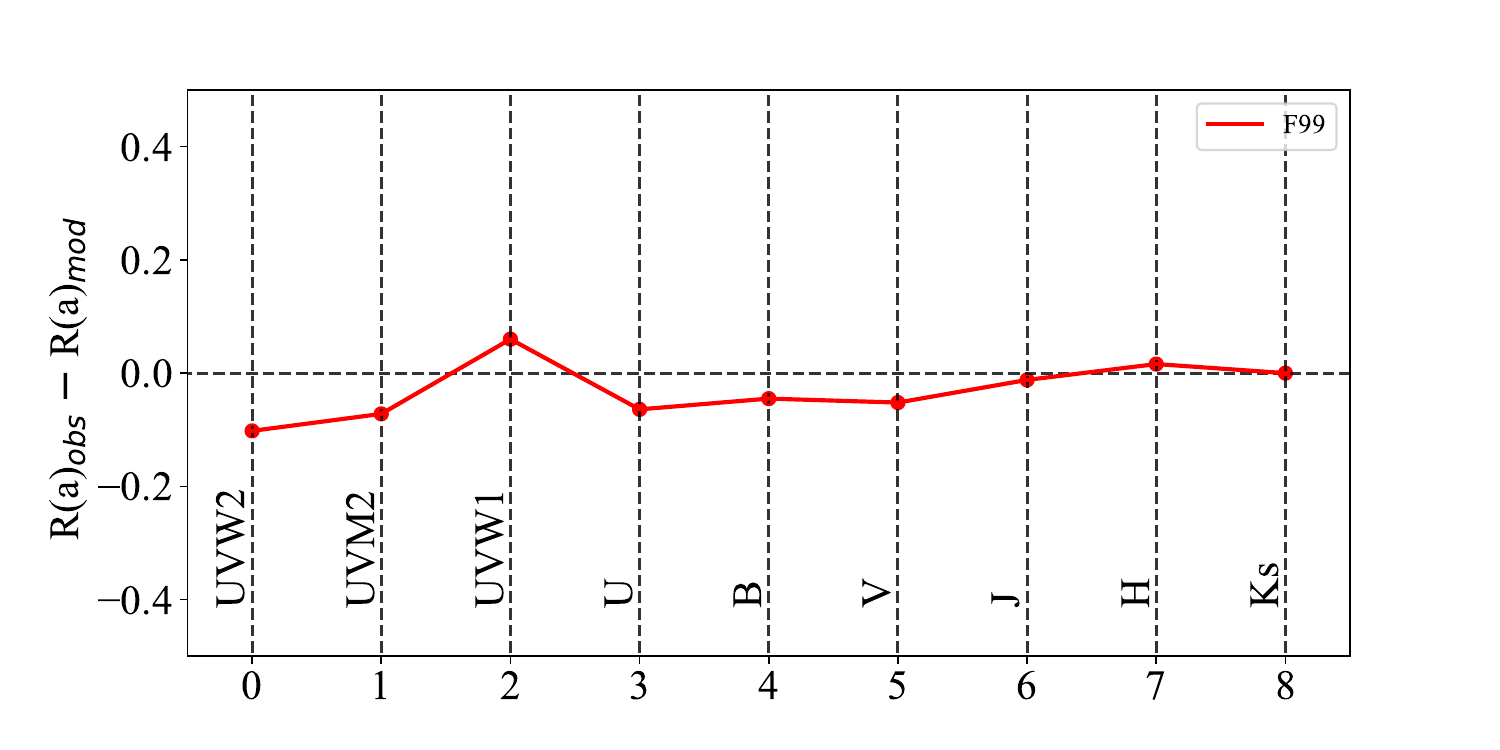}
        \caption{Similar to Figure\,6 but for the comparison of the measured and 
        predicted extinction coefficients.}
        \label{fig7}
    \end{figure}

\subsection{Dependence of reddening coefficients on temperature}

\cite{2023ApJS..264...14Z} showed that empirical reddening and extinction coefficients depend on  stellar temperature and reddening, particularly for blue and broad passbands. 
To investigate the potential dependence of reddening coefficients on temperature, we divided our target sample into low-temperature stars (below 6500\,K) and high-temperature stars (above 6500\,K). For each subsample, their color excess ratios were determined in the same way as for the whole sample. The results are shown in blue and red colors for the low-temperature and high-temperature stars in Figure \ref{fig5}, respectively. Significant discrepancies are found, except for the $B-V$ color.

\begin{figure*}
    \centering
    \vspace{0.2cm}
    \includegraphics[width=14cm]{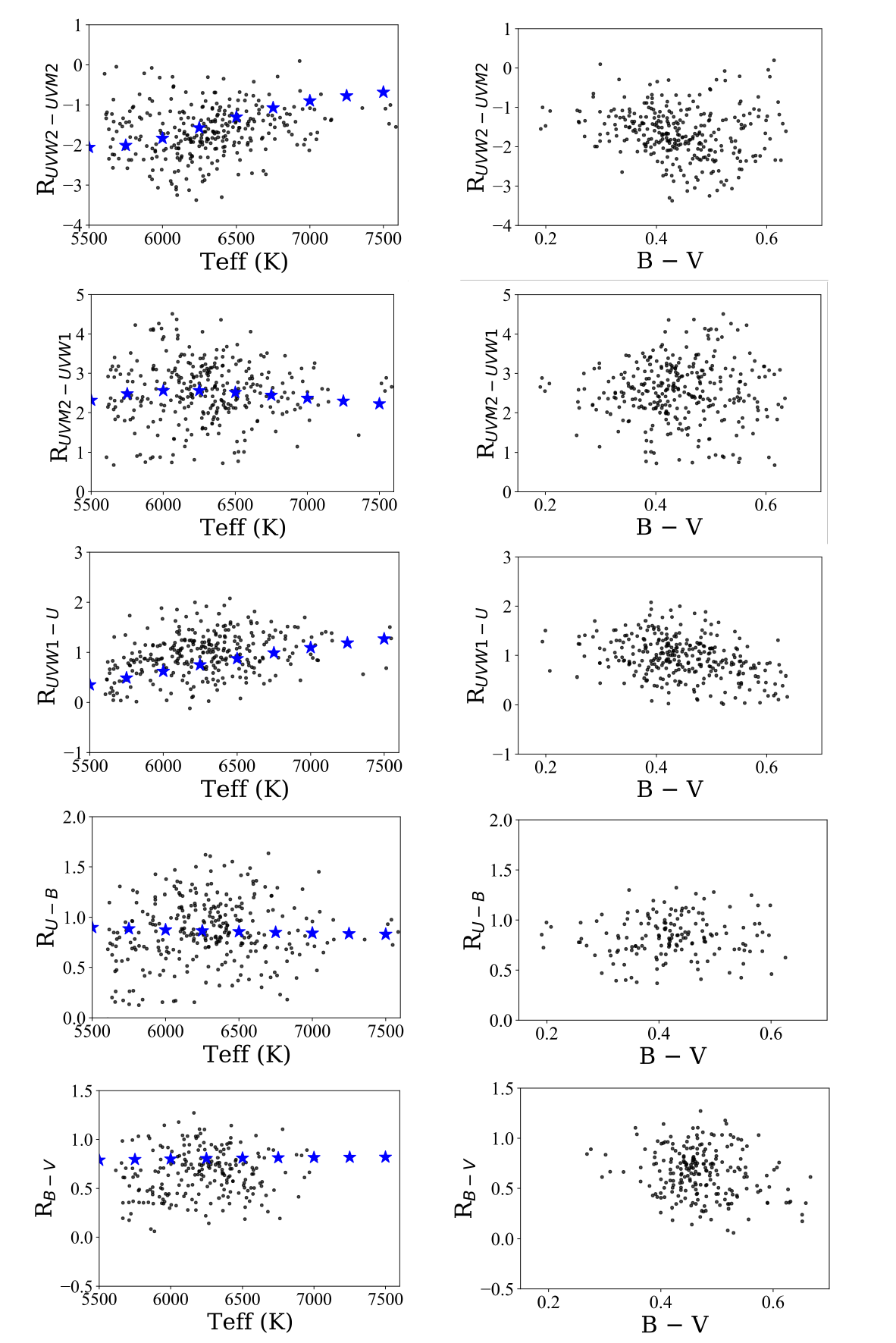}
    \caption{
    Reddening coefficients as a function of stellar temperature (left panels) and $B-V$ (right panels) for individual stars.  Only stars with $E(B-V)$ greater than 0.1 are plotted. Blue asterisks represent predictions of the F99 extinction law of \textit{$R(V)=3.0$} using BOSZ spectra of different temperatures at $E(B-V)=0.2$.  }
    \label{fig8}
\end{figure*}

To better demonstrate the dependence of reddening coefficients on temperature, we plotted reddening coefficients as a function of stellar temperature and $B-V$ color for individual stars in Figure \ref{fig8}. Only stars with $E(B-V)$ greater than 0.1 were used. 
Blue asterisks represent predictions of the F99 extinction law of \textit{$R(V)=3.0$} using BOSZ spectra of different temperatures at $E(B-V)=0.2$. 
It can be seen that the observed and predicted trends agree well. And the dependence on temperature is stronger in the UV bands than in the optical bands. Therefore, the reddening/extinction coefficients in the UV bands should be used with caution.
Limited by the small number of stars in this work, we did not discuss the dependence of reddening coefficients on reddening.

\section{Summary} \label{forth:style}

In this paper, using photometric data from the UVOTSSC and spectroscopic data from LAMOST DR7, we have determined empirical reddening and extinction coefficients for the \textit{Swift}-UVOT passbands with respect to the SFD map. We used the "star-pair" method, which involves selecting pairs of stars with similar intrinsic colors to estimate the amount of reddening. 

We presented the reddening coefficients for the $UVW2-UVM2$, $UVM2-UVW1$, $UVW1-U$, $U-B$, and $B-V$ colors, which are $-1.39$, 2.08, 0.78, 0.72, and 0.84, respectively. The extinction coefficients for the $UVW2$, $UVM2$, $UVW1$, $U$, $B$, and $V$ bands are 5.60, 6.99, 4.91, 4.13, 3.41, and 2.57, respectively. These coefficients are all take the 14\% overestimated into account and recommended to correct the effects of interstellar dust on the observed UVOT colors and magnitudes.

We also compared the results with predictions of Galactic reddening laws and found that the numbers are consistent with predictions by the F99 extinction law of $R(V)$ = 3.0. 
We discussed the temperature-dependent variations of the coefficients, particularly for the UV colors. Cautions should be paid when using the reddening and extinction coefficients for objects
of unusual colors.

\section*{acknowledgments}
The authors sincerely thank the anonymous referee for the helpful comments. 
This work is supported by the National Natural Science Foundation of China through the project NSFC 12222301, 12173007 and 11603002,
the National Key Basic R\&D Program of China via 2019YFA0405503 and Beijing Normal University grant No. 310232102. 
We acknowledge the science research grants from the China Manned Space Project with NO. CMS-CSST-2021-A08 and CMS-CSST-2021-A09.

Guoshoujing Telescope (the Large Sky Area Multi-Object Fiber Spectroscopic Telescope LAMOST) is a National Major Scientific Project built by the Chinese Academy of Sciences. Funding for the project has been provided by the National Development and Reform Commission. LAMOST is operated and managed by the National Astronomical Observatories, Chinese Academy of Sciences.

\section*{DATA AVAILABILITY}
The data underlying this article  are available on line via \url{https://heasarc.gsfc.nasa.gov} and \url{http://dr7.lamost.org/}.

\bibliographystyle{aasjournal}
\bibliography{ref}

\end{document}